\documentclass[10pt,a4paper]{article}
\usepackage[a4paper,margin=0.4in]{geometry}
\usepackage{float}
\usepackage{url,lineno,microtype,subcaption}
\usepackage[english]{babel}
\usepackage{amssymb,amsmath,amsthm,amsfonts}
\usepackage{color,soul}
\usepackage{graphicx}
\usepackage{adjustbox,lipsum} 
\usepackage{esdiff}
\usepackage{physics}
\usepackage[version=3]{mhchem}

\usepackage{siunitx}

\usepackage[superscript,biblabel]{cite}

\usepackage[skip=2pt,font=footnotesize]{caption}

\usepackage{wrapfig}
\usepackage{chemformula} 
\usepackage{mathptmx,cuted}
\usepackage{physics}
\usepackage{sectsty}
\usepackage{fancyhdr}
\usepackage{fnpos}
\usepackage{array}
\usepackage{droidsans}
\usepackage{charter}
\usepackage{epstopdf}
\usepackage{esdiff}
\usepackage{adjustbox,lipsum} 

\newcommand{\pder}[2][]{\frac{\partial#1}{\partial#2}}

\let\oldsqrt\sqrt
\def\sqrt{\mathpalette\DHLhksqrt}
\def\DHLhksqrt#1#2{%
	\setbox0=\hbox{$#1\oldsqrt{#2\,}$}\dimen0=\ht0
	\advance\dimen0-0.2\ht0
	\setbox2=\hbox{\vrule height\ht0 depth -\dimen0}%
	{\box0\lower0.4pt\box2}}
\makeatletter

\usepackage{array}
\newcolumntype{P}[1]{>{\centering\arraybackslash}p{#1}}

\usepackage{url}
\usepackage{xr-hyper}
\usepackage{hyperref}

\title{Dynamics of Moisture Transport in Plant Cuticles:\\
		 The Role of Cellulose}

\author{E.~C.~Tredenick\,$^{1,*}$ and G.~D.~Farquhar\,$^{1}$\\
		\small $^{1}$ARC Centre of Excellence in Translational Photosynthesis, Division of Plant Science,\\ 
		\small Research School of Biology, The Australian National University, Canberra,
		ACT 2601, Australia.\\
	\small $^{*}$e-mail: eloise.tredenick@anu.edu.au
}

	\date{}
\begin{document}
	\onecolumn
	\maketitle	
	
\abstract{Food production needs to increase significantly in 30 years, and water loss from plants may hold one key, especially relevant in a time of climate change. The plant leaf cuticle is the final defence of leaves in drought and at night, and so by understanding water movement in the leaf with mathematical modelling techniques, we can move towards future proofing our crops and native plant ecology. We identify new mechanisms of water movement properties of plant cuticles and utilise this understanding to create a novel mathematical model. We model water sorption in astomatous isolated cuticles, utilising three separate pathways of cellulose, aqueous pores and lipophilic. The results of the model compare well to data both over time and increasing humidity. The sensitivity analysis shows that the grouping of parameters influencing plant species variations has the largest effect on sorption, the parameters influencing cellulose are very influential, and aqueous pores less so but still relevant. Cellulose is important to include in a water transport model for plant cuticles, as it plays a significant role in diffusion and adsorption in the cuticle and the cuticle surfaces.}

\section{Introduction}

Food production may need to increase by 70\% in the next 30 years, in order to feed the world's estimated 9.7 billion people \cite{Hunter2017}, and water loss from plants may hold one key, particularly in a time of changing climate. The plant cuticle is an important layer that covers much of the aerial parts of the plant including the leaves, fruits and non-woody stems. The cuticle is a largely water-proof outer layer that forms the final defence of leaves in drought and at night, where stomata are closed and water loss needs to be kept to a minimum. In the plant cuticle, the mechanisms of transport including adsorption and diffusion of water are, to date, poorly understood. Understanding the transport mechanisms of water can inform a wide range of scenarios from foliar agrochemical application, plant biology and horticulture. 

Several aspects of moisture transport in cuticles have been discovered over the last three decades, including the surface of the cuticle containing salts which may lead to microscopic leaf wetness \cite{Burkhardt2018,Tredenick2020salts}, the water sorption capacity of cellulose \cite{Chamel1991}, aqueous pores\cite{Schreiber2006Agcl,Schonherr2006}, and the influence of the point of deliquescence and hygroscopicity of salts \cite{Tredenick2018}. Moisture sorption in plant cuticles is important to consider and can influence the formation of cracks, leading to issues with food production of fruits such as sweet cherry and apple, and mechanical properties of the plant cuticle \cite{Celino2014, Lopez-Casado2007}. An experimental and theoretical approach has shown that the small fluxes associated with the cuticle are important to include in leaf gas exchange experiments. The work studied several plant species and a drought condition, and found that under certain conditions the cuticular transpiration was an important parameter to include in the calculations\cite{Diego2020submitted}. 

Relative humidity (RH) has a large influence over the weight gain due to water sorption, and penetration of hydrophilic ionic agrochemicals, such as calcium chloride (\ce{CaCl2}), across the cuticle \cite{Schonherr2006,Coret1993,Tredenick2018}, and so the degree of cuticle hydration is important. Humectants and non-ionic surfactants have been shown to influence the hydration of the cuticle \cite{Asmus2016,Coret1993}. Increasing relative humidity can significantly increase penetration of ionic agrochemicals through plant cuticles, and the mechanisms causing this increase are hydration of the cuticle, water adsorption to the aqueous pore walls \cite{Schonherr2006,Luque1995} and water adsorption due to hygroscopic salts that may exist both \textit{in-situ} on the leaf surface \cite{Tredenick2020salts} and included in the spray formulation\cite{Tredenick2018}. Many mechanisms play an important role in water transport, though relative humidity and the mechanisms involved in plant species variations are among the most influential\cite{Tredenick2018}. Within the cuticle, both bound and free water are present\cite{Marechal1996} and we refer to the general process of water travelling into the cuticle as \textit{penetration} or \textit{transport} and the general process where water changes the weight of the cuticle as \textit{sorption}, as there are several specific processes involved in these general processes, such as diffusion (passive Fickian), adsorption (binding) and desorption (unbinding).

Foliar water uptake may play an important ecophysiological role for the plant during drought conditions, though the mechanisms of penetration are not yet fully defined. Foliar water uptake for the plant may be both a cost and a benefit, and a balance is required over the plants lifetime. To date, across the phylogeny, the majority of plant species have been found to have the capacity for foliar water uptake \cite{Dawson2018}. Water may be present on the leaf surface due to rain, dew, high humidity or hygroscopic salts \cite{Tredenick2018}. Foliar water use of a coastal redwood has been characterised and fog suppresses water loss from leaves, ameliorating daily water stress \cite{Burgess2004}. High vapour pressure deficit conditions can reduce photosynthesis and stomatal conductance, while increasing plant water losses through transpiration\cite{Grossiord2020}. Leaf wetting may not only play an important role for plants now, but may be increasingly important in the future due to our changing climate, challenging the plant in novel ways \cite{Dawson2018}. 

The importance of a mechanistic mathematical model has been noted previously in the literature\cite{Tredenick2018,Zabkiewicz2007}, and creating models to describe moisture transport process will improve our understanding of the governing mechanisms. Three mechanistic plant cuticle models have been previously developed\cite{Tredenick2017,Tredenick2018,Tredenick2019Lipo}, focusing on an aqueous solution of hydrophilic ionic (\ce{CaCl2}) agrochemical and lipophilic surfactant penetration into the isolated astomatous plant cuticle, applied as a droplet. The models include mechanisms of ion binding and evaporation with hygroscopic water absorption, along with the ability to vary the active ingredient concentration and type, surfactant formulation, relative humidity, and plant species. Water penetration was included in the model but validation for water diffusion was not conducted, choosing to validate the \ce{CaCl2} penetration data as this was the focus. These validated models form the basis of the current work. 

We aim to characterise the surface properties and transport pathways of water in isolated astomatous cuticles using mechanistic mathematical modelling techniques and to validate the model with well-known experimental data. The model will account for relative humidity, temperature, sorption of water by cellulose, and the ability to model a variety of plant species. The model will be more predictive and less reliant on the need to perform experiments \textit{a priori}. 

The plant cuticle is a structure that is considered the rate-limiting barrier to agrochemical diffusion through plant leaves. It forms a protective layer that is modified by the environment and regulates water loss \cite{Schonherr1989,Kerstiens2010}. The plant cuticle is a lamellate, porous, highly heterogeneous structure that varies between species and individuals in thickness, chemical composition, adaxial and abaxial cuticles, outside and inside surfaces, and abundance and arrangement of structures such as aqueous pores, trichomes, stomata and waxes \cite{Jeffree2006}.

Aqueous pores are dynamic structures within the cuticle that form only in the presence of water \cite{Schonherr2006} and have been visualised across the cuticle surface but are more concentrated in and around the bases of trichomes and stomata \cite{Schreiber2006Agcl,Schlegel2005}. The maximum pore radius varies significantly between plant species, with estimates 0.3~-~2.12~nm\cite{Tredenick2017,Tredenick2019Thesis}. Aqueous pores must not be confused with cracks or permanent and macroscopic pores\cite{brock1983,Schonherr2006}. Hydrophilic ionic agrochemicals penetrate the plant cuticle exclusively through aqueous pores via diffusion \cite{Baur1999diffusion}. Cutin is a major constituent of the polymer matrix within the cuticle, which contains polar polymers. These polar polymers sorb water and swell, giving rise to aqueous pores that traverse the cuticle \cite{Kerstiens2006}. Schreiber~2008\cite{Schreiber20088} notes the chemical nature of these paths can be reasonably speculated on, as plant cuticles are known to have polar functional groups that form the basis of aqueous pores. Non-esterified carboxyl and/or hydroxyl groups of cutin monomers, wax molecules or polar carbohydrates could contribute. Alternatively, if polar carbohydrates extended from epidermal cell walls through the cuticle to the exterior, they could form sites for the diffusion of ionic compounds. 
Lipophilic compounds travel through the cuticle in the lipophilic pathway exclusively, by jumping into voids or defects that arise due to molecular motion by the polymer segments or chains, characterised as a three step process of entering the cuticle via the cuticular lipids, diffusion across the cuticular membrane. The lipophilic pathway is very dependent on temperature and the presence of waxes, when compared to penetration into the aqueous pathway \cite{Schonherr2006,Buchholz2006,Schreiber2001}. 

Water can utilise both the aqueous and lipophilic pathways within the cuticle, as it is a small, uncharged, but polar molecule\cite{Schreiber2006Agcl}. Water molecules within aqueous pores can either diffuse as free molecules \cite{Schonherr2006} or attach to pore walls by adsorption \cite{Luque1995}. The percentage contribution of the aqueous pathway or lipophilic pathway to water penetration in the plant cuticle is currently unknown. However, we can reasonably speculate that this contribution will vary depending on the plant species, environmental conditions and growth conditions. The contribution of each pathway may also be governed by the method of cuticle isolation and chemical treatments, and in the context of agrochemical penetration, pre-treatments and applied droplet chemical composition, including surfactants \cite{Asmus2016,Chamel1989,Chamel1991,Tredenick2019Lipo,Tredenick2020evap}. One study\cite{Schreiber2006Agcl} found that water transport was 2.8 times higher in the aqueous pathway than the lipophilic pathway of astomatous isolated cuticles, and over 15 species of astomatous isolated cuticles, two were incapable of water penetration (ivy and oleander), with significant variation between species. 

The plant cuticle contains polar polysaccharides such as crystalline cellulose, hemicellulose and lignins. Polysaccharides contribute 14\% to 28\% of the dry weight of isolated cuticles \cite{Chamel1991,Lopez-Casado2007}. For example, Chamel~et~al.,~1991\cite{Chamel1991} found that the contribution to the total water sorption in a cuticle is attributed 67\% to polysaccharides, 32\% to cutin and 1\% to waxes. Although the weight of polysaccharides in the cuticle is relatively small, they contribute to the majority of water sorption and hydrogen bonds \cite{Chamel1991,Marechal1996}. Plant cuticle sorption curves have been found to be strongly influenced by the removal of polysaccharides, especially at high relative humidity. The components of polar polysaccharides in tomato fruit cuticles is 25 - 33\% cellulose, 26 - 30\% pectin and 20 - 24\% hemicellulose \cite{Lopez-Casado2007}. Polysaccharides are an important constituent of all cuticles studied to date and cellulose is present in leaf and fruit cuticles, especially on the inside surface, but also throughout the cuticle \cite{Guzman2014, Lopez-Casado2007,Jeffree2006}. Cellulose has been visualised with gold labelling in enzymatically isolated and intact leaf cuticles of eucalyptus, poplar, and pear, and was observed in the whole cuticle of these species except eucalyptus, where it was mainly in the cuticular layer \cite{Guzman2014}.

\begin{figure} [h!]
	\centering
	\includegraphics[width=0.38\textheight]{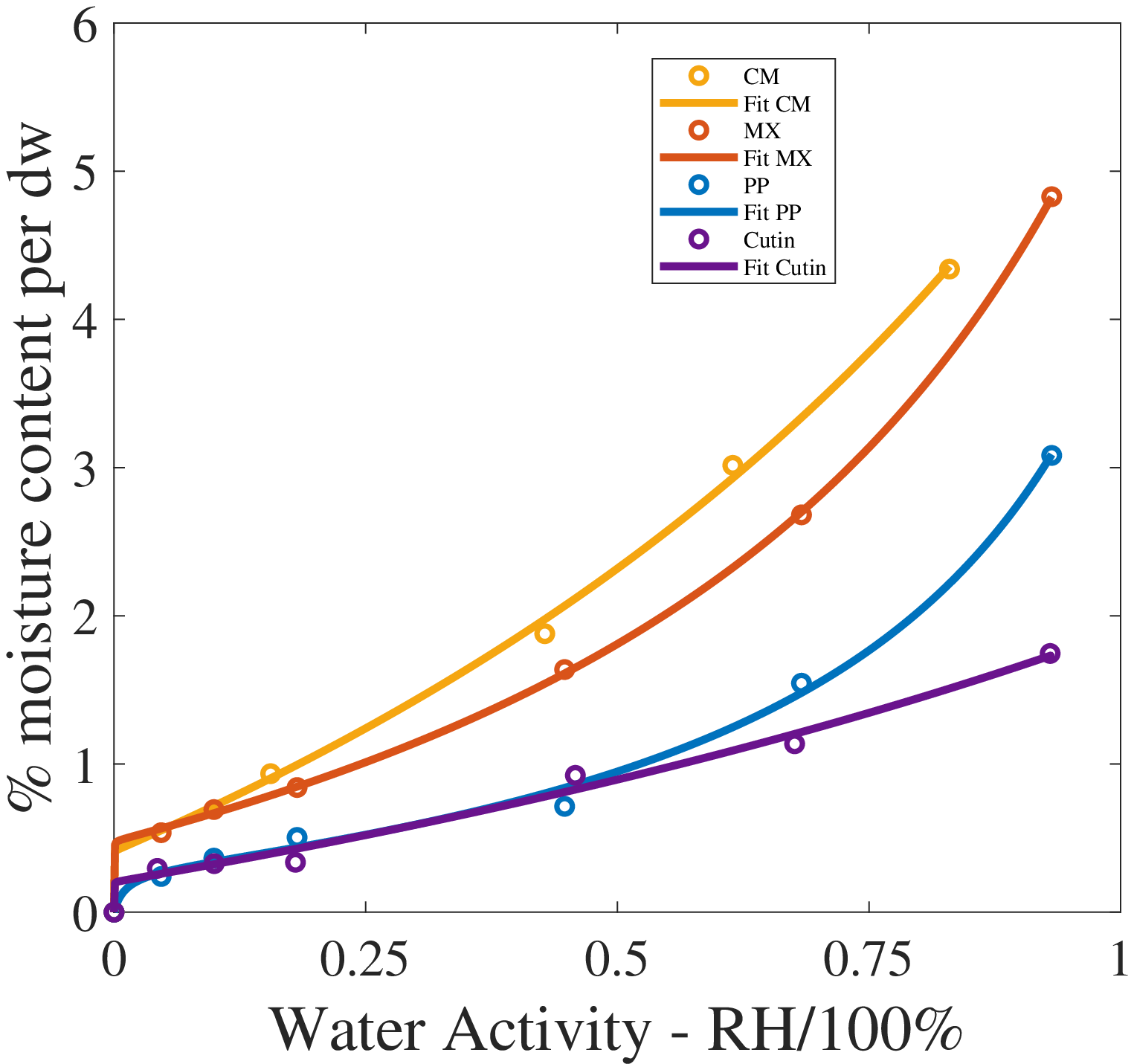}
	\caption{Moisture sorption curves for the cuticle membrane (CM), polymer matrix (MX) membranes, cutin (free of polysaccharides), and polar polysaccharides (PP) only, which is MX minus cutin. The MX and CM curves are very similar. When PPs are removed (referred to as cutin) in purple, the large nonlinear increase in sorption at high humidities is no longer present and the sorption increases linearly. The large nonlinear increase in the PP curve can clearly be seen at high humidities, indicating PPs are the reason for this trend. We fitted\cite{Luque1995} the data\cite{Chamel1991} using $y = a \ b \ x/(1+a \ x) + c \ d \ x/(1-c \ x)$, where $y$ is the percentage moisture adsorption per dry weight and $x$ is the water activity or RH$/100\%$, with parameters shown in Table \ref{table:Fits}. }
	\label{fig:ChamelMX}
\end{figure}

The water sorption of isolated cuticles increases with relative humidity, and there is also an additional significant increase at high humidities, as shown in Figs.~\ref{fig:RHSA} and \ref{fig:ChamelMX}. When the isolated cuticle is treated to create a hydrolysed/cutin sample, that is free of polar polysaccharides including crystalline cellulose\cite{Chamel1991}, the increase at high humidities is much less pronounced and has a linear trend. Therefore, we can conclude that the polar polysaccharides (cellulose) in the cuticle causes the large increase in sorption at high humidities. As there is more research around cellulose than hemicellulose and pectin in the cuticle, and it has been found to be the major polysaccharide\cite{Dominguez1999}, we will henceforth refer to polar polysaccharides simply as cellulose.

Binding or adsorption of ionic compounds to the surfaces of isolated cuticles is possible, and the capacity of binding is more on the inside and less on the outside cuticle surface \cite{Yamada1964}. In some cases, the difference is 12 to 68 times, being higher on the inside surface. The timescale and amount of ions that penetrated was also different depending on the application cuticle side, for several plant species and compounds. The total number of ions travelling from the outside to the inside surface (when applied to the outside) was always larger than the inside to the outside (when applied to the inside). After 40 hours, the penetration had not yet reached steady-state\cite{Yamada1964}. Extending this mechanism to the whole leaf, the cuticle is more likely to gain ionic compounds from applied agrochemicals and aerosols deposited from the atmosphere and is less likely to lose ions from inside the plant leaf interior. Permeability to water also has the same directional dependence as ionic compounds\cite{Schieferstein1959}. Cellulose was found in surfaces of cuticles in two independent works\cite{Schieferstein1959,Herediaguerrero2014}, generally having a larger proportion on the inner surface. The presence of cellulose and directional dependence implies water can bind to cellulose on the cuticle surface. When applying this to the whole leaf, water penetration entering from outside is more likely and the cuticle is less likely to lose water from the inside of the leaf. It is important to include a mechanism in a water sorption or penetration model where water is trapped or bound to the cuticle surface with some degree of asymmetry.

There is still the question of how and why this high sorption takes place at high humidity, as shown in Fig.~\ref{fig:ChamelMX}. It is well known that cellulose is highly hygroscopic and can sorb 30\% of water per dry weight, and polar polysaccharides, isolated from cuticles, can sorb even more at 49\% \cite{Dominguez1999}. Regarding polysaccharides in cuticles, very little work has been done on composition, molecular characteristics, physical or chemical behaviour \cite{Lopez-Casado2007}. Water molecules can adopt many interaction configurations of surrounding molecules due to the two hydrogen bond acceptor and donor sites. Water also provides many opportunities to be inserted into a polymer with the ability to play a multitude of roles \cite{Marechal1996}. Many studies using different experimental techniques\cite{Guo2017, Marechal1996, Marechal1997, ChamiKhazraji2013, Guo2018, Chen2018}, have found three types of hydrogen bonds; cellulose to other cellulose groups, water to cellulose groups and water to other water molecules\cite{Chen2018} (where water is attached to cellulose, noted to be tetrahedral in shape\cite{Guo2017}). Above 80\%RH, the share of water bonds contributed to water to other water is the most abundant kind of bond\cite{Guo2017,Chen2018}. The bond of water to cellulose is stronger than the bond of water to other waters upon desorption, and the water to water bond is easier to break\cite{Chen2018}. We surmise that at high humidity, above around 55\%RH \cite{Dominguez1999}, the cellulose to hydrogen bonds are mostly full, so water to water bonds form, and water sorption increases significantly at high humidity. We note that more research needs to be done, with particular focus on cellulose at high humidity as the mechanisms driving increased sorption at high humidities are not fully understood, and why water to water bonds form more frequently at high humidities than other kinds of bonds. 

\section{Results}
\subsection{Model Framework}
We describe a novel comprehensive mechanistic model for isolated cuticle water transport. The model takes the form of a nonlinear, one-dimensional diffusion model, including partial differential equations. We will briefly describe the modelling formulation, that is based on the authors previous works \cite{Tredenick2017,Tredenick2018,Tredenick2019Lipo} and we refer the reader to these works for a full description of auxiliary equations (\ref{diffh2oA}), (\ref{radius}), (\ref{lang}), (\ref{GammaS}) and (\ref{Beta}).

Aqueous pores take up water and their properties vary between plant species, with relative humidity, and are less influenced by temperature and the presence of waxes than the lipophilic pathway \cite{Schonherr2006,Buchholz2006}. Cutin takes up water and this is the likely location of aqueous pores, while cellulose takes up more water than cutin\cite{Chamel1991}.  The sorption ability of cellulose is higher than other cuticle materials and it is not homogeneously distributed\cite{Guzman2014,Schieferstein1959,Herediaguerrero2014}. We model cellulose as a separate pathway from the aqueous pores, which is reasonable due to the results shown in Fig.~\ref{fig:ChamelMX}, as the results of cutin (aqueous pores) and polar polysaccharides (cellulose) are different.

We account for water, that can travel in three pathways; aqueous pores (A), the lipophilic pathway (L) and cellulose (C). These three pathways are governed by different mechanisms, so must be modelled separately. The component of water changes primarily along the cuticle membrane thickness, $z$ ($0 \leq z \leq b$). Moisture inside the cuticle is considered liquid water, which is a reasonable assumption based on experimental data\cite{Marechal1996,Schreiber2001}. This model will make novel additions to a simple diffusion model by accommodating the unique pathways of the cuticle. We incorporate the important governing mechanisms of swelling of the aqueous pores, climatic conditions such as temperature and relative humidity that affect the pore swelling and adsorption, parameters that account for differences in plant species including porosity and tortuosity, pore density and radius, cuticle thickness and binding to the cuticle surface due to cellulose. The model, as described in equations (\ref{h2o})--(\ref{ICrA}), (\ref{GammaS})--(\ref{percdw}), parameters in Table \ref{Parameters} and constants in Table \ref{Constants}, for the diffusion and adsorption of water into the three pathways of the cuticle, is as follows: 
\begin{align}
	\left( \varepsilon_{\scriptscriptstyle \text{C}} +  \varepsilon_{\scriptscriptstyle \text{L}} \right) \ \pder[ c  ]{t} +   \pder[\left(\varepsilon_{\scriptscriptstyle \text{D}} c  \right)]{t}    & = \left( \varepsilon_{\scriptscriptstyle \text{C}} \ D_{\scriptscriptstyle \text{C}} +\varepsilon_{\scriptscriptstyle \text{L}} \ D_{\scriptscriptstyle \text{L}} \right) \  \pdv[2]{ c }{z}
	+\pder[{}]{z} \left( D_{\scriptscriptstyle \text{A}}  \pder[({\varepsilon_{\scriptscriptstyle \text{D}} c   })]{z} \right) -  \frac{2}{r_{\scriptscriptstyle \text{A}}   }\left(1-\varepsilon_{\scriptscriptstyle \text{D}}\right) \pder[{\Gamma_{\scriptscriptstyle \text{A}}      }]{t}  - \rho_{\scriptscriptstyle \text{C}} \ \left(1 - \varepsilon_{\scriptscriptstyle \text{C}}\right) \pder[{\Gamma_{\scriptscriptstyle \text{C}}      }]{t} , \ \ 0<z<b, \ t>0, \label{h2o}  
\end{align}
where $c$ is the concentration of liquid water in the cuticle, $\varepsilon_{\scriptscriptstyle \text{D,C,L}}$ are the porosities, $t$ is time, $D_{\scriptscriptstyle \text{A,C,L}}$ are the diffusivity functions, $z$ is the thickness of the cuticle, $r_{\scriptscriptstyle \text{A}}$ is the aqueous pore radius, $\Gamma_{\scriptscriptstyle \text{A,C}}$ is the concentration of adsorbed water in the aqueous pores and cellulose, $\rho_{\scriptscriptstyle \text{C}}$ is the density of cellulose fibres, and the subscripts A, D, C and L are aqueous pores (A is the entire aqueous pore and D is the aqueous pore available for diffusion), cellulose and lipophilic pathway. The parameters that change in space and time include $c(z,t),\  D_{\scriptscriptstyle \text{A}}(z,t), \ r_{\scriptscriptstyle \text{A}}(z,t), \ \varepsilon_{\scriptscriptstyle \text{D}}(z,t), \ \Gamma_{\scriptscriptstyle \text{A,C}}(z,t)$.

The experimental setup\cite{Chamel1991}, which is utilised to design the model including the domain and initial conditions, consists of astomatous isolated tomato fruit cuticles, placed in a range of relative humidities. The cuticles are initially dry, but may contain a small amount of water, and the relative humidity is increased in steps. The cuticles are weighed and the increase in weight, above the dry weight is recorded, in two different experiments, one at a near-steady-state at 6 hours over a large range of humidities, and the other over 10 minutes and a small selection of humidities. Moisture can enter the cuticle at both the outside and inside cuticle surface. At the boundary,  water travels into the cuticle, and adsorbs to the cuticle surface due to the presence of cellulose on the surface, where it is trapped and no longer available for diffusion, but can increase the weight of the cuticle. Water then enters the cuticle and can travel down the three pathways. In the aqueous pores, the free water molecules can travel through the cuticle via passive Fickian diffusion, or be adsorbed in a monolayer to the aqueous pore walls, causing them to swell. The aqueous pores swell due to increasing water content from increasing relative humidity and increase in size until they reach the maximum pore radius \cite{Tredenick2017}. In the lipophilic pathway, water can diffuse through the cuticle and is influenced by temperature change. For cellulose, water can adsorb to the cellulose on the cuticle surface, then free water molecules can diffuse through the cuticle, or adsorb to cellulose fibres. In equation (\ref{h2o}), the last two terms govern adsorption in the cuticle aqueous pores and cellulose. The term governing adsorption in a monolayer is formulated by including the surface area of a pore over the pore volume including porosity, and the cellulose formulation instead utilises the density of cellulose and the includes multilayer adsorption. Both free and adsorbed water increases the weight of the cuticle and are required to calculate $\Delta w$. These adsorption isotherms are described in more detail around equations (\ref{lang})--(\ref{eq:GAB}).

The cuticle surface mechanisms involved in sorption are significant. The cuticle experimental setup is such that the relative humidity is applied at both the outside and inside boundary conditions. The outside and inside (where the inside is touching the interior of the leaf if the leaf was intact) cuticle surfaces, at $z=0$ and $z=b$, are governed by the following boundary conditions:
\begin{align}
-   \left. \left[ \left(\varepsilon_{\scriptscriptstyle \text{C}} \ D_{\scriptscriptstyle \text{C}} + \varepsilon_{\scriptscriptstyle \text{L}} \ D_{\scriptscriptstyle \text{L}} \right) \ \pdv[]{ c }{z} + D_{\scriptscriptstyle \text{A}} \ \pder[({\varepsilon_{\scriptscriptstyle \text{D}} c   })]{z} \right]  \right\vert_{z=0}& =  - h \ \left(c_\infty - c(0,t) \left( 1 - \dfrac{ c(0,t)}  {c^{\scriptscriptstyle \text{pure}}_{\scriptscriptstyle \text{H$_2$O}}} \right) \right)  -   k_1 \ c(0,t), \quad  t>0,   \label{BCz0}\\
-  \left. \left[ \left(\varepsilon_{\scriptscriptstyle \text{C}} \ D_{\scriptscriptstyle \text{C}} + \varepsilon_{\scriptscriptstyle \text{L}} \ D_{\scriptscriptstyle \text{L}} \right)  \ \pdv[]{ c }{z} + D_{\scriptscriptstyle \text{A}} \ \pder[({\varepsilon_{\scriptscriptstyle \text{D}} c   })]{z} \right] \right\vert_{z=b} &=  h \ \left(c_\infty - c(b,t) \left( 1 - \dfrac{ c(b,t)}  {c^{\scriptscriptstyle \text{pure}}_{\scriptscriptstyle \text{H$_2$O}}} \right) \right) -  k_2 \ c(b,t), \quad  t>0,   \label{BCzb}
\end{align}
where $h$ is the moisture transfer coefficient for moisture going into the cuticle, $c_\infty$ is the atmospheric vapour concentration of air far from cuticle as a function of temperature, $c^{\scriptscriptstyle \text{pure}}_{\scriptscriptstyle \text{H$_2$O}}$ is the concentration of water as a function of temperature, and $k_1$ and $k_2$ are the rate constants for water binding to cellulose on the outside and inside cuticle surfaces. Equations (\ref{BCz0}) and (\ref{BCzb}) are Robin type boundary conditions, analogous to Newton's law of heating, and shows that the flux of moisture across the boundary is proportional to the difference between the atmospheric vapour concentration and moisture levels on the surface of the cuticle. We have extended the condition analogous to Newton's law of heating by including a logistic growth formulation that prevents the concentration of water exceeding that of pure water, necessary due to the values of the constants in the equation. Water can adsorb to the cellulose in the cuticle \cite{Yamada1964,Schieferstein1959,Herediaguerrero2014}, so it is important to include in the model. The final term of equations (\ref{BCz0}) and (\ref{BCzb}) describes the adsorption to cellulose on the cuticle surface, and increases proportional to the concentration of moisture of the surface. The parameters $k_1$ and $k_2$, the rate constants for water adsorbing to cellulose on the cuticle surfaces, is a function of relative humidity and the amount of cellulose present, as described in equation (\ref{WcCellulose}) and Table \ref{Constants}. The outer and inner surfaces are similar, except they contain different amounts of cellulose, so different amounts of water can bind. Additional equations included in the mathematical model are described in Sections \ref{methodsv2} and \ref{extra model22}.

\subsection{Model Validation}\label{valid}

\begin{figure} [h!]
	\centering
	\includegraphics[width=0.5\textheight,keepaspectratio]{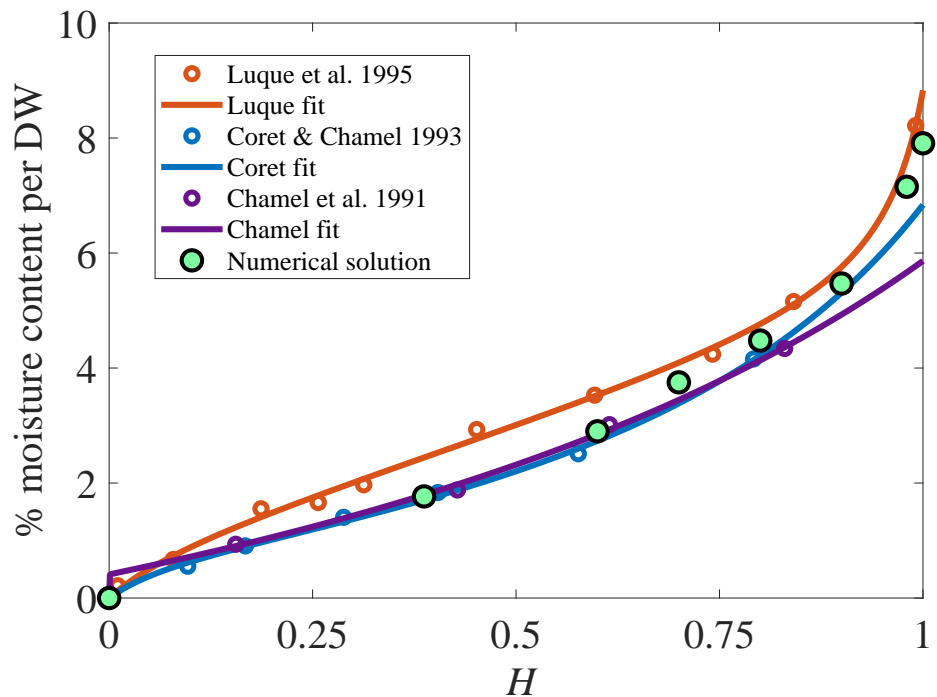}
	\caption{Validation plot compared to experimental data \cite{Luque1995,Coret1993,Chamel1991} on isolated astomatous tomato fruit cuticles. The total experimental time in the data\cite{Chamel1991} and model is 6 hours for each point, and experimental data are shown as open circles. Relative humidity, $H$, is shown as a fraction on the x-axis and the percentage moisture content on the y-axis is calculated with equation \ref{percdw}. The green dots are the numerical solution results of the cuticle model, and the solid curves are simple fits described around Table \ref{table:Fits2}.}
	\label{fig:RHSA}
\end{figure}

\begin{figure*} [h!]
	\centering
	\includegraphics[width=0.75\textheight,keepaspectratio]{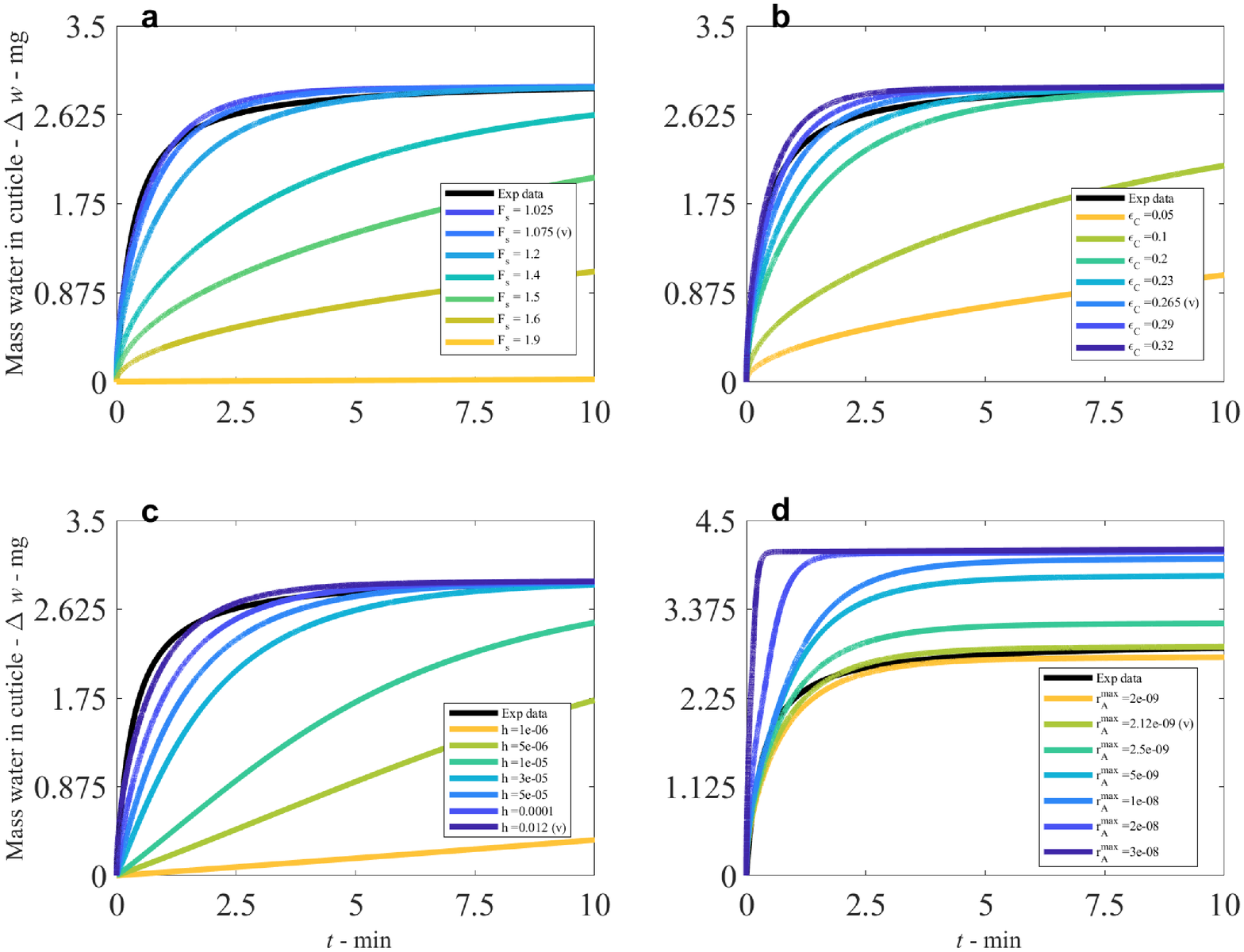}
	\caption{Sensitivity analysis of the model, compared to the results for validation (v) and the experimental data (Exp data in black) as a reference point, all performed at 60\%RH. All model parameters are kept constant except the sensitivity parameter. (a) shows the sensitivity of $F_s$, the fractal scaling dimension, analogous to the tortuosity, (b) shows $\varepsilon_{\scriptscriptstyle \text{C}}$, which is the porosity of cellulose, (c) shows $h$, the moisture transfer coefficient at the boundaries, and (d) shows $r_{\scriptscriptstyle \text{A}}^{\scriptscriptstyle \text{max}}$, the maximum radius of aqueous pores. Note for all the subfigures, the relationship is directly proportional, except for $F_s$, which has an inverse relationship.}
	\label{fig:allSA}
\end{figure*}

The model, as described in equations (\ref{h2o})--(\ref{ICrA}), (\ref{GammaS})--(\ref{percdw}), is solved numerically. We then validate the model solution against experimental data\cite{Chamel1991}. We validate the model over two sets of results, one with a range of relative humidities, and the second with a selection of humidities and time. Fig.~\ref{fig:RHSA} shows the results of the model after 6 hours of experimental time (green dots), compared to the experimental data, over a range of relative humidities. Due to the variable nature of cuticles, we have compared the model results to three experiments on isolated tomato fruit cuticles with a similar experimental setup (open circles)\cite{Luque1995,Coret1993,Chamel1991}, and fit (continuous curves) their data with a simple equation, as described in Table \ref{table:Fits2}. In Fig.~\ref{fig:RHSA}, we can see the model compares well to all three experiments over the range of humidities. In particular, the model increases at a large degree (above a linear trend), at very high humidities, at 90\%RH and 99\%RH, which is the desired result. This trend at high humidities is largely the result of water attaching to the cellulose on the outside and inside cuticle surfaces, discussed later around Fig.~\ref{fig:allSA}. In isolated cuticle experiments, studying penetration of ionic active ingredients, relative humidity also increases penetration\cite{Schonherr2000,Schonherr2002}, and therefore the model compares well to the well-established data.

\begin{figure}[h!]
	\centering
	\includegraphics[width=0.3\textheight,keepaspectratio]{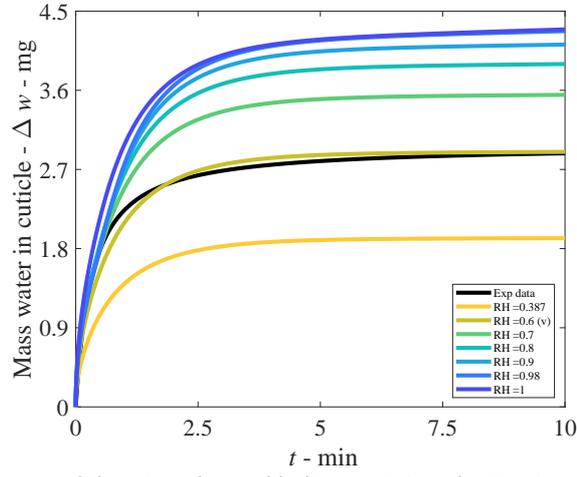}
	\caption{Validation plot compared to experimental data (Exp data in black) at 57\%RH. The 60\%RH model solution (orange) can be seen responding closely to the experiment ((v) represents the solution used for validation). The total experimental time here is 10 minutes. We reproduce the experimental data in black, as described around equation (\ref{chamelfits1}).}
	\label{fig:valid2SA}
\end{figure}

\begin{figure} [h!]
	\centering
	\includegraphics[width=0.3\textheight,keepaspectratio]{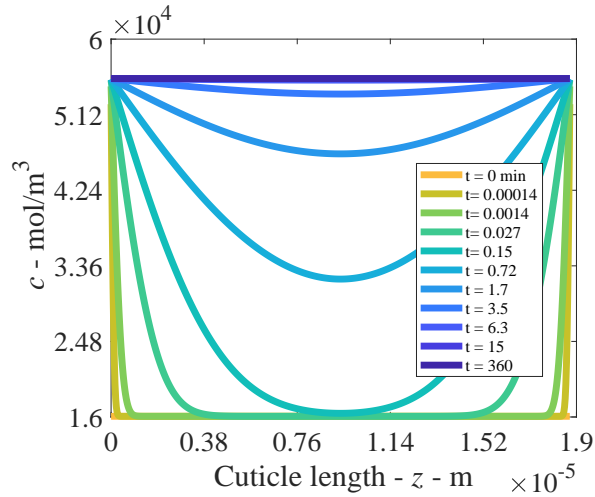}
	\caption{Plant cuticle diffusion model results at a single relative humidity of 60\%RH, focusing on the concentration of free water molecules over 6 hours. Over the length of the cuticle, results are shown at a selection of times. The initial condition is shown in orange and the boundary conditions for the outside and inside cuticle surfaces are located at cuticle length 0 and $1.87\times 10^{-5}$~m, respectively.}
	\label{fig:conc}
\end{figure}

In Fig.~\ref{fig:valid2SA}, we see the model compares well to the experimental data, at 60\%RH over time, and the change in dry weight increases proportional to relative humidity. The initial rapid increase is largely produced by cellulose and aqueous pores, with less influence from the lipophilic pathway, discussed around Fig.~\ref{fig:allSA}. In Fig.~\ref{fig:conc}, we see the concentration plot of free water molecules at a selection of times with the length of the cuticle. Initially, (orange) the water concentration is low and constant through the cuticle. Then, as time increases, the concentration at the boundaries increases and then gradually diffuses into the cuticle. At late times, around 12 minutes, the cuticle is close to the concentration of pure water. This figure only shows free water concentration and not bound water as the other figures do. Due to the asymmetric boundary conditions, the concentration of water at the two boundaries is similar but not identical. The model validation is reasonable and compares well to the experimental data of four separate experiments, from three\cite{Luque1995,Coret1993,Chamel1991} water sorption experiments.

We conduct a sensitivity analysis of the model, based on the validation result (v) in Fig.~\ref{fig:valid2SA} at 60\%RH, keeping all the parameters the same, then utilising the one-factor-at-a-time method to determine the models parameter sensitivities, as shown in Fig.~\ref{fig:allSA}. A selection of sensitivities is discussed below. 
Fig.~\ref{fig:RHSA} and \ref{fig:valid2SA} have already displayed the sensitivity of relative humidity and been validated with the experiential data. In Fig.~\ref{fig:allSA} (a), we see the results of changing $F_s$, the fractal scaling dimension, analogous to the tortuosity, of the three pathways. The increase in weight of the cuticle is very sensitive to the tortuosity of the pathways, as this modifies the diffusivities. This is the only inverse relationship produced, with increasing $F_s$ meaning a more tortuous and longer pathway that is more difficult to cross, slowing and limiting the diffusivity and diffusion of the three pathways. When applying this result to isolated cuticle experiments, the tortuosity of aqueous pores will differ significantly between plant species cuticles, due to plant anatomy such as lamellate structures, thickness, orientation of pores and plant age\cite{Santier1992, Schreiber2006Agcl,Jeffree2006}. Therefore the significant effect of $F_s$, indicates that the variation of plant species itself, has a significant effect on water transport, and this is seen experimentally\cite{Schreiber2006Agcl}.

In Fig.~\ref{fig:allSA} (b), the sensitivity of $\varepsilon_{\scriptscriptstyle \text{C}}$ is shown, the porosity of cellulose. This parameter has a strong direct relationship to mass increase, and shows similar (but direct not inverse) trends to $F_s$, as these parameters modify the effective diffusivity of cellulose (equation (\ref{diffC})). A larger porosity value implies more pores for diffusion, so this trend is as expected. The strong effect of modifying a parameter that only applies to this one pathway, implies cellulose contributes significantly to mass increase of the cuticle. However, when changing porosity, the mass result remains unchanged at 10 minutes for most porosities chosen, and this is because when one pathway is limited or removed from the model, the other two pathways can compensate within the model. In Fig.~\ref{fig:allSA} (c), we see the sensitivity to $h$, the moisture transfer coefficient at the boundaries. The parameter $h$ has a large effect as it modifies the three pathways, and changes the time when the water enters the cuticle but not the maximum value at late times. Fig.~\ref{fig:allSA} (d) shows $r_{\scriptscriptstyle \text{A}}^{\scriptscriptstyle \text{max}}$, the maximum radius of the aqueous pores. A direct relationship exists, and increasing the maximum pore size promotes faster diffusion and adsorption into the pores. The maximum pore radius has a somewhat significant effect over the change in mass increase of the cuticle. The implication for this effect of changing this parameter for aqueous pores is that aqueous pores do contribute somewhat to the overall mass increase, but significant contributions also are produced by cellulose. 

The other parameters studied (results not shown), including the rate coefficient for cellulose binding on the boundaries, $k$, the porosity of the lipophilic pathway, $\varepsilon_{\scriptscriptstyle \text{L}}$, and temperature, $T$. The rate constant for binding at the boundary conditions due to cellulose, $k$, has very little effect on mass increase in the short term, but a large effect over longer times. This indicates that binding is over a long timescale and too slow compared to the diffusion timescale, and that cellulose binding contributes significantly over long periods to the results shown in Fig.~\ref{fig:RHSA}. Two timescales are evident here, a short diffusion and adsorption timescale governed largely by cellulose and aqueous pores diffusion and adsorption inside the cuticle, and a longer timescale governed by adsorption onto the cellulose at the boundaries, and these two timescales are reflected by both experiments and mathematical models\cite{Chamel1991,Tredenick2018,Tredenick2019Lipo}. The mechanism of binding on the surface is crucial to include in a water sorption model so we can validate the results in Fig.~\ref{fig:RHSA}, and must be included, as it was in previous cuticle models\cite{Tredenick2017,Tredenick2018,Tredenick2019Lipo}. The porosity of the lipophilic pathway, $\varepsilon_{\scriptscriptstyle \text{L}}$, has a weak effect on mass increase, which implies the lipophilic pathway does contribute to mass increase somewhat, but not to a significant degree in this study. This agrees with results from Chamel~et~al.,~1991\cite{Chamel1991}, where wax extraction only had a small effect and results found elsewhere\cite{Schreiber2006Agcl} that water travels mostly down the aqueous pathway and less down the lipophilic pathway. We note some studies\cite{Asmus2016,Coret1993} have used other techniques, chemicals, pre-treatments and adjuvants from Chamel~et~al.,~1991\cite{Chamel1991}, that may increase the contribution of the lipophilic pathway to water penetration. There is a small effect of changing temperature, $T$, and temperature changes 6 parameters relating to the properties of water and the lipophilic pathway, as shown in Table \ref{Constants}. The lipophilic pathway is influenced by temperature and aqueous pores less so\cite{Schonherr2006,Buchholz2006} and Chamel~et~al.,1991\cite{Chamel1991} find only a weak relationship between temperature and mass increase of the cuticles studied, therefore these results are in keeping with the well-established literature\cite{Chamel1991,Schonherr2006,Buchholz2006}. 

The parameters that are highly influential over sorption of the cuticle are related to variations in plant species, including $F_s, b, \ r_{\scriptscriptstyle \text{A}}^{\scriptscriptstyle \text{max}}$ and $\eta_{\scriptscriptstyle \text{A}}$. The parameters related to cellulose, including $\varepsilon_{\scriptscriptstyle \text{C}}$, $k$, and the parameters related to $k$ and $\Gamma_{\scriptscriptstyle \text{C}}$, are also highly influential over sorption in the model, and this aligns with experimental data, as cellulose sorbs significant amounts of water in cuticles\cite{Chamel1991,Dominguez1999}. Relative humidity also has a large effect on mass increase. Our results from the sensitivity analysis are feasible. Considering all the results from the sensitivity analysis, we find that cellulose and the aqueous pores contribute to mass increase of the cuticle in the shorter timescale (10 minutes), while the lipophilic pathway contributes to a small degree. In the longer timescale (6 hours), diffusion in cellulose and binding to the cellulose at the boundaries drives the mass increase.

\section{Discussion}\label{sec:Discussion}

The model can simulate water sorption in the cuticle with any relative humidity, temperature, content of cellulose on the surface, cuticle thickness, aqueous pore radius and pore density. The model can be theoretically applied to any plant species by modifying the constants $b, \ DW, \ c_{\scriptscriptstyle \text{min}}, \ F_{\scriptscriptstyle \text{s}}, \ n , \ r_{\scriptscriptstyle \text{A}}^{\scriptscriptstyle \text{max}}, \  \rho_{\scriptscriptstyle \text{C}}, \ \eta_{\scriptscriptstyle \text{A}}$. The model does not include any mechanisms that are specific to any one plant species, and future work could include validating the results for other plant species, noting that the trends for other plant species are very similar in Chamel~et~al.,~1991\cite{Chamel1991}. The model only requires one set of data to be trained on, and many parameters are available in the literature, as described in Table \ref{Constants}.

When extending this model to an attached plant leaf (though not considering stomatal interactions), the previously held assumption was that the interior was 100\% humidity or equivalent to liquid water. With these conditions, water loss to the environment may occur through the cuticle from the interior of the plant, when the environment is at low relative humidities. For transport in the other direction (inwards), new data (S.~C.~Wong and G.~D.~Farquhar, unpublished data), suggests the humidity at the cuticle interior is less than 100\%RH in some conditions, and this could assist with the modelling, when this work is expanded to the attached plant leaf. Considering transport in astomatous cuticles attached to leaves, at isothermal conditions, if the interior boundary condition or initial condition of the cuticle is less than 100\%RH, water will enter the cuticle and subsequently the plant from the exterior, as water goes down the concentration gradient based on passive diffusion.

The mechanism of water adsorption and transport in cellulose could lead to a further understanding of the transport of hydrophilic uncharged agrochemicals though cuticle. Uncharged methyl glucose has been found to have influencing mechanisms that differed from hydrophilic ionic or lipophilic compounds, indicating an alternative hydrophilic pathway within the cuticle. 60\% of methyl glucose diffused across the aqueous pathway, while 40\% used an alternative pathway \cite{Shi2005, Schonherr2006}. Perhaps this alternate pathway could involve cellulose. We note that if the isolated cuticle was being heated and subsequently drying out, there will be vapour present inside the cuticle, as was found experimentally\cite{Marechal1996}. To model the transport of vapour and liquid in a heated cuticle, the model could be adapted, as shown elsewhere for other plant materials \cite{Fadai2017}.

Describing the aqueous pathway as a pore, is one way to represent this pathway \cite{Schreiber2005,Schonherr2006}. This is a suitable way to include these pathways in a mathematical model with a continuous modelling approach, where these pores or voids may be very small or grow larger and fill with water and change size in space and time with water content. We note that this pathway has also been described as a dynamic aqueous continuum \cite{Beyer2005,Fernandez2017physico}, that is only continuous through the cuticle when polar functional groups are clustered \cite{Schonherr2006}. Here we develop this further by separately modelling aqueous pores and cellulose, along with the lipophilic pathway, totalling three pathways, which is a more comprehensive approach. Modelling three separate pathways is important as their governing mechanisms are different. With further mechanistic research, a more defined view of the aqueous continuum can develop to further understand water transport in the cuticle and leaf. We note some research has been done in the past 5 years on the cuticle and cellulose, but more research is needed for comprehensive insights into water transport in the cuticle, and this is a key reason for employing mathematical modelling here.

To conclude, we describe a comprehensive novel model for isolated cuticle water transport to move towards a better understanding of the transport of water within an attached intact leaf. This is the first comprehensive mechanistic model to simulate and validate moisture transport in cuticles. We bring together both new and old knowledge, and create a more mechanistic perspective of water transport in the plant cuticle. This paper has highlighted the importance of including mechanisms influencing water transport in cellulose in a cuticle model. The model validates well and the sensitivities align with the well-established literature. By understanding the governing mechanisms, we can move towards improved modelling of whole attached plant leaves, agrochemical formulation development and application, understanding water loss from plants in a time of changing climate conditions, and transport of liquid and gaseous water applied to the cuticle.

\section{Methods}
\subsection{Model Description} \label{methodsv2}
The diffusivities for water travelling in the three pathways are as follows, where the formulations are similar and the lipophilic pathway is governed by temperature:
\begin{align}
	D_{\scriptscriptstyle \text{A}}(z,t)&  = D_{\scriptscriptstyle \text{H$_2$O}}^{\text{bulk}} \ \varepsilon_{\scriptscriptstyle \text{D}}\ ^{ \left( \frac{F_{\text{s}}} {2-F_{\text{s}} } \right)}, \quad   0<z<b, \ t>0,   \label{diffh2oA}\\
	D_{\scriptscriptstyle \text{C}} & = D_{\scriptscriptstyle \text{H$_2$O}}^{\text{bulk}} \ \varepsilon_{\scriptscriptstyle \text{C}} \ ^{ \left( \frac{F_{\text{s}}} {2-F_{\text{s}} } \right)} , \label{diffC}\\
	D_{\scriptscriptstyle \text{L}} & = D_{\scriptscriptstyle \text{H$_2$O}}^{\text{bulk}} \ e ^{\left( \frac{-E}{R\ T} \right)} \ \varepsilon_{\scriptscriptstyle \text{L}} \ ^{ \left( \frac{F_{\text{s}}} {2-F_{\text{s}} } \right)}  , \label{diffh2oL}
\end{align}
where $D_{\scriptscriptstyle \text{H$_2$O}}^{\text{bulk}} $ is the self/bulk diffusion coefficient of water as a function of temperature, $F_{\text{s}}$ is the fractal scaling dimension, analogous to tortuosity, and $E$, $R$, and $T$ are constants described in Table \ref{Constants}. The formulation for the changing aqueous pore radius is as follows, with more details provided elsewhere \cite{Tredenick2017}:
\begin{align}
r_{\scriptscriptstyle \text{A}}(z,t) &= r_{\scriptscriptstyle \text{H$_2$O}} \left( 1 + \left( \sin \left(\left(\Gamma_{\scriptscriptstyle \text{A}} \ r_{\scriptscriptstyle \text{H$_2$O}}^2 \ N \right)^{-1} \right) \right) ^{-1} \right), \quad  0<z<b, \ t>0.  \label{radius}
\end{align} 
The following equations are utilised to model adsorption:
\begin{align}
	\Gamma_{\scriptscriptstyle \text{A}}(z,t) &  = \frac{ \Gamma_{\scriptscriptstyle \text{S}} \ \beta \  c }{1 + \beta  \ c },\quad \quad \quad \quad  0<z<b, \ t>0, \label{lang} \\
	\Gamma_{\scriptscriptstyle \text{C}}(z,t) &  = \frac{ \Gamma_{\scriptscriptstyle \text{SC}} \ \beta_{\scriptscriptstyle \text{C}} \ K \  c }{\left(c^{\scriptscriptstyle \text{pure}}_{\scriptscriptstyle \text{H$_2$O}}       - K \ c \right) \left( 1 + K \ \left(\beta_{\scriptscriptstyle \text{C}} - 1 \right)  \ \dfrac{c}{c^{\scriptscriptstyle \text{pure}}_{\scriptscriptstyle \text{H$_2$O}}} \right) },\quad \quad  0<z<b, \ t>0, \label{eq:GAB} 
\end{align}
where $\Gamma_{\scriptscriptstyle \text{A}}$ is the concentration of water adsorbed per unit area in aqueous pores, $\Gamma_{\scriptscriptstyle \text{C}}$ is the concentration of water adsorbed in cellulose,
and the constants are described in Table \ref{Constants}. Cellulose adsorption is modelled with a water adsorption isotherm model, which will contribute significantly at high humidity, due to the formation of multilayers of water. The adsorption for cellulose is modelled with the Guggenheim, Anderson, and De Boer (GAB) isotherm (based on the BET isotherm), $\Gamma_{\scriptscriptstyle \text{C}}$, as shown in equation (\ref{eq:GAB}) \cite{Brunauer1938,Guggenheim1966,DeBoer1953,Anderson1946}. The GAB isotherm describes water adsorption as a monolayer, which can then form multilayer at high humidities. Water adsorption to the aqueous pores is modelled with a Langmuir isotherm, as shown in equation (\ref{lang}), which describes adsorption as a monolayer and further details are discussed in previous works\cite{Tredenick2017}. The radius for the aqueous pore for diffusion of water molecules is slightly smaller than the entire pore ($r_{\scriptscriptstyle \text{A}}$), and water molecules are arranged in a monolayer or closed Steiner chain on the pore surface. We account for this by formulating porosity as follows:
\begin{align}
\varepsilon_{\scriptscriptstyle \text{D}}(z,t)   & = \pi \left( \left( r_{\scriptscriptstyle  \text{A}} - 2 \ r_{\scriptscriptstyle \text{H$_2$O}}  \right) \left( \sqrt{\eta_{\scriptscriptstyle  \text{A}}}+   \dfrac{1}{L}     \right) \right ) ^{2} , \  0<z<b, \ t>0,   \label{eps_4}
\end{align}
where $\varepsilon_{\scriptscriptstyle \text{D}}$ is the porosity of the aqueous pores for diffusion, $r_{\scriptscriptstyle \text{H$_2$O}}$ is the Van der Waals radius of a water molecule,
$\eta_{\scriptscriptstyle  \text{A}}$ is the density of aqueous pores in cuticle and $L$ is the control volume length. The term $r_{\scriptscriptstyle  \text{A}} - 2 \ r_{\scriptscriptstyle \text{H$_2$O}}$ accounts for the smaller pore radius, and more details are available elsewhere on the formulation\cite{Tredenick2017}. The initial conditions of the model dictate a dry cuticle that has a small amount of water present, and the smallest aqueous pore radius where free water can exist, is described as follows:
\begin{align}
c(z,0) & = c_{\scriptscriptstyle \text{min}}, \quad    0 \leq z \leq b,   \label{ICh20}\\
r_{\scriptscriptstyle \text{A}} (z,0) &= 3 \ r_{\scriptscriptstyle \text{H$_2$O}},   \quad    0<z<b,   \label{ICrA}
\end{align}
where $c_{\scriptscriptstyle \text{min}}$ is the concentration of water in a somewhat dry cuticle as a function of relative humidity and a radius of three waters is the minimum to allow the transport of water molecules.

\begin{table*} [h!]
	\centering
	\caption[Model parameters]{Model parameters.}	 	\label{Parameters}
	\footnotesize 	
	\begin{tabular}{    p{2cm}  p{7cm}  p{2cm}  } 
		\hline
		{ Parameter} & {Definition} & 		{Value and Units} \\ \hline

		{  $c(z,t)$}        &  {  Concentration of water}       & 
		{  mol$/$m$^3$}                 \\

		{  $D_{\scriptscriptstyle \text{A}}(z,t)$, $D_{\scriptscriptstyle \text{L}}$, $D_{\scriptscriptstyle \text{C}}$} & {Diffusivity of water in the aqueous pores, lipophilic pathway and cellulose pathway} & 
		{  m$^2/$s}  \\ 

		{  $r_{\scriptscriptstyle \text{A}}(z,t)$}        &  {  Radius of aqueous pores}       & 
		{ m  }               \\

		{ $t$}        &  { Time}       & {s}                \\

		{ $z$} & {Thickness} & 	{ m}    \\

		$ \varepsilon_{\scriptscriptstyle \text{D}}(z,t) $ & Aqueous pathway porosity for diffusion  &   \\

		{ $\Gamma_{\scriptscriptstyle \text{A}}(z,t)$}        &  {Concentration of water adsorbed per unit area in the aqueous pathway}       & 
		{  mol$/$m$^2$}           {}          \\

		{$\Gamma_{\scriptscriptstyle \text{C}}(z,t)$}        &  {Concentration of water adsorbed in cellulose}       & 
		{  mol$/$kg}           {}          \\

		$\Delta w(t)$ &  Weight increase of cuticle above dry weight  &  mg  \\

		\hline
	\end{tabular}
\end{table*}

The model, as described in equations (\ref{h2o})--(\ref{ICrA}), (\ref{GammaS})--(\ref{percdw}), is solved numerically, similar to previous models\cite{Tredenick2018,Tredenick2019Lipo}. We use `ode15i' within MATLAB\textsuperscript{\textregistered}, with a finite volume method, averaging of the diffusivity function, $D_{\scriptscriptstyle \text{A}}(z,t)$, for aqueous pores at the control volume faces, and discretise the model's partial differential equations with second order central differences to approximate the spatial derivatives, with evenly distributed nodes. The fitted parameters as described in Table \ref{Constants}, are found by fitting to the data\cite{Chamel1991} at 60\%RH, then keeping all the parameters the same for all other simulations and only changing the humidity to match the experimental humidity.

The time scale of moisture sorption is important as it informs the development of models. Sorption from 5 plant species starts to level out at around 5 to 10~minutes at humidities lower than 70\% but high humidities are not measured, and samples are also measured for 6 hours minimum but the maximum is not stated, at each humidity and this data set is not measured with time\cite{Chamel1991}. For moisture sorption in cellulose, 20\%RH took around 1 hour, while sorption at 95\%RH took 5.6~hours to level out \cite{Guo2018}. An experimental time of 2-3~days is common with gravimetric techniques \cite{Tredenick2020salts}, agrochemical penetration experiments typically occur over 2-3 days \cite{Tredenick2017}, while with a radioactive water sorption technique in cuticles, 20-28~hours was used \cite{Schreiber2001}. Therefore, we can conclude that for the Chamel~et~al.,~1991\cite{Chamel1991} data after 6 hours, the sorption has not necessarily reached equilibrium in a mathematical sense, and data measured with time over 3 days is necessary to establish the true point of equilibrium. We note it is important to define the usage of the term equilibrium, and here we define it as the time at which the rate of change or time derivative is zero. However, in an experimental sense, the experiment will often end sooner at a near-steady-state, due to a variety of reasons. The timescale of aqueous pore swelling is also an important consideration, but there has not yet been a definitive study with a range of relative humidities and plant species to define this timescale and more work needs to be done. Here, aqueous pores swell close to their maximum at 9 minutes, but may take longer to reach equilibrium.

\section*{Acknowledgements}
The authors acknowledge funding provided by the Australian Research Council Centre of Excellence for Translational Photosynthesis (CE1401000015). The authors wish to thank Nicole Pontarin for the insight and advice provided on the manuscript. 

\section*{Author contributions}
ECT is responsible for article writing, model creation and adaptation, computational code creation, results, analysis, article editing and revision. 
GDF is responsible for model adaptation, analysis, article editing and revision. 

\section*{Conflict of interest statement}
The authors declare that the research was conducted in the absence of any commercial or financial relationships that could be construed as a potential conflict of interest.

\section*{Additional information}
Supplementary Information accompanies this paper.

\clearpage
\section*{Supplementary Materials}

\subsection{Experimental Data Fitting}

\begin{table*} [h!]
	\centering 
	\caption[Fit parameters]{Fit parameters for Fig.~\ref{fig:ChamelMX}.}	 	\label{table:Fits}
	\small	
	\begin{tabular}{    p{1.3cm}  p{1.6cm}  p{1.6cm}  p{1.6cm} p{1.6cm}} 
		\hline
		{Parameter} & {CM} & {MX}& {PP}& {Cutin}\\ \hline 
		{a } &  {$1.44\times 10^{5}$}  & {$1.8\times 10^{4}$} &  {91.64} &  {$2.04\times 10^{4}$}   \\
		{b}  &  {0.41}  & {0.48 } &  {0.28} &  {0.2}   \\
		{c } &  {0.46}  & {0.67} &  {0.78} &  {0.31}   \\
		{d } &  {6.36}  & {2.67} &  {1.06} &  { 3.75}   \\
		{R$^2$ \%} &  {99.9}  & {99.9} &  {99.6} &  {99.0}   \\
		\hline
	\end{tabular}
\end{table*}

The experimental data over a time of 10 minutes, as shown in Fig.~\ref{fig:valid2SA}, was shown in the original text\cite{Chamel1991} as uncorrected for  water sorption by the vessel or basket holding the sample that is also being weighted and can also adsorb water. We have corrected for the vessel here by fitting two Langmuir isotherms (see equation (\ref{lang})), then the sample minus the vessel, to produce the weight of water in the cuticle, $\Delta w(t)$, at 57\%RH, with an R-squared value better than 99\%, as follows:
\begin{align}
	\Delta w(t) & =\dfrac{3.758 \times 3.437 \times t}{3.437 \times t+1} - \dfrac{6.056 \times 0.7808 \times t}{6.056  \times t+1},\label{chamelfits1}
\end{align}
where $t$ here is time in minutes. The original work\cite{Chamel1991} did not provide error bars, and the two sets of data points, for the sample data not corrected for by the vessel and the data for the vessel, were not conducted at matching relative humidities, hence we are only able to produce a fitted curve, not individual data points.

\begin{table*} [h!]
	\centering 
	\caption[Fit parameters]{We find fitted parameters for solid lines in Fig.~\ref{fig:RHSA}, using $y = a \ b \ H/(1+a \ H) + c \ d \ H/(1-c \ H)+e\ H$, where $y$ is the percentage moisture adsorption per dry weight\cite{Luque1995} and $H$ is the water activity or RH$/100\%$.
	}	 	\label{table:Fits2}
	\small	
	\begin{tabular}{    p{1.3cm}  p{1.5cm}  p{2.7cm}  p{1.4cm} } 
		\hline
		{Parameter} & {Luque \cite{Luque1995}} & {Coret and Chamel \cite{Coret1993} }& {Chamel \cite{Chamel1991}}\\ \hline 
		{a } &  {8.55 }  & { 10.76} &  { $3.89\times 10^{4}$}    \\
		{b}  &  {0.99 }  & { 0.79 } &  {0.41 }  \\
		{c } &  { 0.95}  & {  0.66 } &  { 0.46}    \\
		{d } &  {0.22 }  & { 3.09} &  { 6.35}   \\
		{e } &  {4.02 }  & {0 } &  {0 }    \\
		{R$^2$ \%} &  {99.83}  & {99.82} &  {99.85}    \\
		\hline
	\end{tabular}
\end{table*}

\section{Additional Model Description} \label{extra model22}
Here we describe additional equations, included in the model, as shown in equations (\ref{GammaS})--(\ref{percdw}) and the modelling constants in Table \ref{Constants}. The parameters $\Gamma_{\scriptscriptstyle \text{S}}$ and $\beta$ are calculated as follows:
\begin{align}
	\Gamma_{\scriptscriptstyle \text{S}} &= \left( r_{\scriptscriptstyle \text{H$_2$O}} ^2 \ N \arcsin( \dfrac {r_{\scriptscriptstyle \text{H$_2$O}} } {	r_{\scriptscriptstyle \text{A}}^{\text{max}} \ H - r_{\scriptscriptstyle \text{H$_2$O}}})    \right) ^{-1} , \label{GammaS}\\
	\beta& = \frac{\arcsin( \dfrac {r_{\scriptscriptstyle \text{H$_2$O}} } {	r_{\scriptscriptstyle \text{A}}^{\text{max}} \ H - r_{\scriptscriptstyle \text{H$_2$O}}}) 
	}{ c_{\scriptscriptstyle \text{H$_2$O}}^{\text{pure}}  \left(    \arcsin( \dfrac {r_{\scriptscriptstyle \text{H$_2$O}} } {	r_{\scriptscriptstyle \text{A}}^{\text{max}} \ H^2 - r_{\scriptscriptstyle \text{H$_2$O}}})  - \arcsin( \dfrac {r_{\scriptscriptstyle \text{H$_2$O}} } {	r_{\scriptscriptstyle \text{A}}^{\text{max}} \ H - r_{\scriptscriptstyle \text{H$_2$O}}}) \right)} . \label{Beta} 
\end{align}
The constant $\Gamma_{\scriptscriptstyle \text{S}}$, described in equation (\ref{GammaS}), is described elsewhere\cite{Tredenick2017}, and here the maximum pore radius, $r_{\scriptscriptstyle \text{A}}^{\text{max}}$, is limited by relative humidity, $H$. The constant $\beta$, as described in equation (\ref{Beta}), is formulated utilising equations (\ref{lang}) and (\ref{GammaS}), simplifying and rearranging. To calculate the binding of water to cellulose, $k_2$, as described in Table \ref{Constants}, on the surface of the cuticle as a function of humidity, $H$, we utilise the following equation to find $W_C$,
\begin{equation}
	W_C = K_1 \ \left( \dfrac{K_2 \ H}{1 + K_2 \ H} + \dfrac{K_3 \ H}{1 - K_3 \ H}\right) ,\label{WcCellulose}
\end{equation}
where $W_C$ is the weight of water adsorbed per gram of dry solid as a fraction, $H$ is the relative humidity as a fraction or water activity, $K_1$ is the number of strong binding sites and equal to 0.05, $K_2$ is the attraction of these sites and equal to 7.43, and $K_3$ is related to the water activity of the solid and equal to 0.907 \cite{Dominguez1999}. All parameters are dimensionless and $a_{\scriptscriptstyle \text{surf}}$ scales the outside surface, as there is less cellulose\cite{Schieferstein1959,Herediaguerrero2014} on the outside surface (see Table \ref{Constants}).

\subsection{Conversion of concentration to weight}
To convert the final solution from a concentration to a weight in mg, including the adsorbed water, the following equations are applied. The initial condition is removed from the solution, as the initial condition is equivalent to the dry weight, to produce 
\begin{equation}
	c^w (z,t) = c(z,t) - c_{\scriptscriptstyle min} ,\label{c_final}
\end{equation}
where $c^w$ is the concentration of water without the initial condition, $c$ is the concentration of free water and solution to the model, and $c_{\scriptscriptstyle \text{min}}$ is the initial condition.

The experimental data are given as the difference between the wet weight, at a given relative humidity, and dry weight over time; therefore to convert the concentration to the total weight in mg, the following equation is utilised. The first integral is considered over space (resulting in a solution at each point in time), and the second is a cumulative integral over time, and the first three terms are ions in the cuticle, adsorbed in aqueous pores and adsorbed in cellulose, and the cumulative integral term is the ions adsorbed to cellulose at the two cuticle surfaces,
\begin{equation}
	\Delta w (t)  = M_{\scriptscriptstyle \text{w}}  \ f  \ n \left[ \int_{0}^{b} \left[  A_{\scriptscriptstyle \text{CM}} \ c^w (z,t)  +  2 \ \pi \ r_{\scriptscriptstyle \text{A}} (z,t)  \ \Gamma_{\scriptscriptstyle \text{A}}( c^w (z,t) )   + \dfrac{f \ A_{\scriptscriptstyle \text{CM}} \ \rho_{\scriptscriptstyle \text{C}} \ \Gamma_{\scriptscriptstyle \text{C}}( c^w (z,t) ) }
	{\Gamma_{\scriptscriptstyle \text{SC}} \ M_{\scriptscriptstyle \text{w}}}  \right] dz +  A_{\scriptscriptstyle \text{CM}} \int_{0}^{t_{\scriptscriptstyle \text{final}}} \left[ k_1 \ c^w (0,t)  + k_2 \ c^w (b,t) \right] dt \right],\label{deltamg}
\end{equation}
where $\Delta w(t)$ is the weight increase over dry weight in mg, $M_{\scriptscriptstyle \text{w}}$ is the molecular weight of water, $f$ is the value $10^{3}$ and converts g to mg or kg to g, $n$ is the total number of cuticle discs used in the experiment, $t_{\scriptscriptstyle \text{final}}$ is the final experimental time and is 10 minutes here, $A_{\scriptscriptstyle \text{CM}}$ is the area of one cuticle disc, $z$ is the thickness of the cuticle, $\Gamma_{\scriptscriptstyle \text{A}}$ is the adsorption of water to the aqueous pores, $\Gamma_{\scriptscriptstyle \text{C}}$ is the adsorption of water to cellulose, $\Gamma_{\scriptscriptstyle \text{SC}}$ is the saturated concentration of water adsorbed in cellulose, $r_{\scriptscriptstyle \text{A}}$ is the aqueous pore radius, $t$ is time and $k_1$ and $k_2$ are the rate constants for binding to cellulose on the cuticle surfaces. The formulation of the area for the moles adsorbed per aqueous pore, using $\Gamma_{\scriptscriptstyle \text{A}}$, is based on the circumference of the aqueous pore that is circular in cross-section.

To convert $\Delta w$ to a percentage increase over the dry weight at the end time, we utilise the following:
\begin{equation}
	\% \text{moisture content per DW} = \dfrac{   \Delta w (t_{\scriptscriptstyle \text{final}}) \ 100\% }  {  DW \ n     }      ,\label{percdw}
\end{equation}
where $\% \text{moisture content per DW}$ is the percentage gain of water content over the dry weight, $DW$ is the total dry weight and $t_{\scriptscriptstyle \text{final}}$ is 6 hours here. The resultant change in weight, $\Delta w (t)$, is a vector and can be seen in Fig.~\ref{fig:valid2SA}, while the $\%$ moisture content per DW is a scalar at each RH and can be seen in Fig.~\ref{fig:RHSA}.

\begin{table*} [h!]
	\caption[Model constants]{Model constants.}	 	\label{Constants}
	\footnotesize 	
	\begin{tabular}{    p{1cm}  p{6.5cm}  p{3cm}  p{7.5cm} } 
		\hline
		{ } & {Definition} & 		{Value and Units} &  {Comments} \\ \hline 
		
		{  $A_{\scriptscriptstyle \text{CM}}$}        &  { Cuticle area of one surface}       & 
		{  $7.85\times 10^{-5}$~m$^2$}          &  {}          \\  
		
		{  $b$}        &  {  Thickness of cuticle}       & 
		{  $1.87\times 10^{-5}$~m}          &  {  \cite{Chamel1991}}          \\

		{  $c_{\scriptscriptstyle \text{min}}$}        &  {Initial concentration of water in a relatively dry cuticle, as a function of RH}       & 
		{mol$/$m$^3$}          & {For [0, 0.387, 0.6, 0.7, 0.8, 0.9, 0.98, 0.999]RH, $c_{\scriptscriptstyle \text{min}}=$[0, 4515, 7168, 4515, 2596.1, 1140.9, 208, 300]}\\

		{  $c^{\scriptscriptstyle \text{pure}}_{\scriptscriptstyle \text{H$_2$O}}$}  &  { Pure water concentration as a function of temperature }       &
		{mol$/$m$^3$}          &  {$\rho_{\scriptscriptstyle \text{H$_2$O}} /  M_{\scriptscriptstyle \text{w}}$  }          \\  
		
		{  $c_\infty$}        &  {Atmospheric vapour concentration of air far from cuticle, as a function of temperature, $T$, in K }       & 
		{ mol$/$m$^3$ }          &  {$ P_{\text{v}} \ H / \left( R \ T \right)$}          \\

		{  $D_{\scriptscriptstyle \text{H$_2$O}}^{\text{bulk}}$}        &  {  Self/bulk diffusion coefficient of water as a function of temperature }       & 
		{m$^2/$s}          &  {For temperatures [20, 25, 30, 35, 40, 45, 50]$\,^{\circ}\mathrm{C}$, [2.022, 2.296, 2.59, 2.919, 3.24, 3.575, 3.968]$\times 10^{-9}$~m$^2/$s  \cite{Mills1973,Easteal1989} }          \\

		{DW}        &  {Total dry weight of isolated tomato fruit cuticle for experiment with humidity}       & 
		{137.7~mg }          &  { Calculated based on experimental data. This is the equivalent of using 85 individual cuticle discs for the experiment conducted over 6 hours } \\

		{  $E$} & {Activation energy for diffusion in the lipophilic pathway} & 	 {    $2.64\times 10^{4}$~Pa~m$^3/$mol  } &  { For waxy isolated tomato fruit cuticles \cite{Knoche1994}  } \\

		{  $F_{\scriptscriptstyle \text{s}}$} & {  Fractal scaling dimension} & 	 {   1.075} &  {Chosen to agree with experimental data} \\ 	
		
		{$h$} & {Moisture transfer coefficient} &  { 0.012~m$/$s} & {Chosen to agree with experimental data }\\

		{$H$} & {  Relative humidity as a fraction} & 		{ } &  {$H = RH/100\% =  p / p_{\scriptscriptstyle \text{sat}}  $, $p$ is the partial pressure of water vapour, $p_{\scriptscriptstyle \text{sat}}$ is the pressure of saturated water vapour} \\ 
		
		{  $k$}        &  {Rate constant for water binding to cellulose on the cuticle surface  }       & 
		{$6.5\times 10^{-9}$~m$/$s}          &  { Chosen to agree with experimental data }          \\ 
		
		{  $k_1$}        &  { Rate constant for water binding to cellulose on the outside cuticle surface   }       & 
		{m$/$s  }          &  { $k_1 = a_{\scriptscriptstyle \text{surf}} \ k_2$, where $ a_{\scriptscriptstyle \text{surf}} = 0.1$}          \\ 
		
		{  $k_2$}        &  {Rate constant for water binding to cellulose on the inside cuticle surface  }       & 
		{m$/$s   }          &  { $k_2 = W_C \ k$, where $W_C$ is defined around equation (\ref{WcCellulose}) \cite{Dominguez1999} as a function of relative humidity }          \\

		{  $K$}        &  { GAB isotherm constant }       & 
		{  0.46}          &  { \cite{Bedane2016} }          \\

		{  $L$}        &  {  Control volume length}       & 
		{  1~m}          &  {  }          \\ 
		
		{  $M_{\scriptscriptstyle \text{w}}$} & {  Molecular weight water} & 
		{  $18.015$~g$/$mol} &  {  } \\ 
		
		{  $N$}        &  {  Avogadro constant}       & 
		{   $6.022\times 10^{23}$~mol$^{-1}$}          &  { }          \\ 
		
		{ $n$}        &  {Number of tomato cuticle discs used in timed experiment over 10 minutes }       & 
		{ 1 }          &  { Calculated based on experimental data \cite{Chamel1991}}          \\

		{  $P_{\scriptscriptstyle \text{v}}$}        &  {  Saturated water vapour pressure in air as a function of temperature in $^{\circ}\mathrm{C}$, $T_C = 20\,^{\circ}\mathrm{C}$}       & 
		{Pa}      &  {  $611.21 \exp[(18.678-T_C/234.84)(T_C/(273.15+T_C)]$  }          \\  
		
		{  $R$}        &  {  Gas constant}       & 
		{  8.31~Pa$\cdot$m$^3/$K$/$mol}          &  { }\\ 	
		
		{  $r_{\scriptscriptstyle \text{H$_2$O}}$}        &  {Van der Waals radius of a water molecule}       & 
		{  $1.5\times 10^{-10}$~m}          &  {  }          \\  	
		
		{  $r_{\scriptscriptstyle \text{A}}^{\scriptscriptstyle \text{max}}$}        &  {  Maximum radius of aqueous pores}       & 
		{  $2.12\times 10^{-9}$~m}          &  {  For tomato fruit cuticle, \cite{Schreiber2009}}          \\

		{  $T$} & {Temperature} & 	{ K } & \\ 
		
		{  $\beta$} & { Langmuir parameter} & 
		{m$^3/$mol} &  {  Equilibrium parameter of adsorbed water} \\

		{  $\beta_{\scriptscriptstyle \text{C}}$} & { GAB parameter} & 
		{4} &  {  Equilibrium parameter of adsorbed water for GAB isotherm \cite{Bedane2016}} \\

		$ \varepsilon_{\scriptscriptstyle \text{L}} $ & Lipophilic pathway porosity  & 0.03 & \cite{Tredenick2019Lipo} \\
		
		$ \varepsilon_{\scriptscriptstyle \text{C}} $ & Cellulose porosity  & 0.265 & Chosen to agree with experimental data \\
		
		{  $\eta_{\scriptscriptstyle \text{A}}$}        &  { Density of aqueous pores in cuticle}       & 	  {	$2.18\times 10^{15}$~m$^{-2}$}         &  {\cite{Tredenick2018} }          \\

		{  $\Gamma_{\scriptscriptstyle \text{S}}$}        &  {Langmuir saturation constant}       & 
		{ mol$/$m$^2$}          &  {  $0<\Gamma_{\scriptscriptstyle \text{A}}<\Gamma_{\scriptscriptstyle \text{S}}$, saturation concentration of water adsorbed per unit area in the aqueous pathway }          \\

		{  $\Gamma_{\scriptscriptstyle \text{SC}}$}        &  {GAB constant}       & 
		{ 6663~mol$/$kg}                  &  {Monolayer concentration of water adsorbed per volume cellulose \cite{Bedane2016}}          \\ 
		
		{  $\rho_{\scriptscriptstyle \text{H$_2$O}}$}        &  {  Liquid density of water as a function of temperature}       & 
		{  g$/$m$^3$}   &  {$	 (999.848 + 6.338\times 10^{-2}  T_C - 8.524\times 10^{-3}  T_C^2   + 6.943\times 10^{-5}    T_C^3 -3.821\times 10^{-7}   T_C^4)   1000$	\cite{Jones1992its} }          \\

		{$\rho_{\scriptscriptstyle \text{C}}$} & {Density of cellulose fibres } &  { 1450~kg$/$m$^3$} &  { \cite{Bedane2016}  } \\

		\hline
	\end{tabular}
\end{table*}

\end{document}